\documentclass[sigconf]{acmart}

\usepackage{svg}
\usepackage{booktabs} 
\usepackage{enumitem}

\AtBeginDocument{%
  }

\copyrightyear{2023}
\acmYear{2023}
\setcopyright{rightsretained}
\acmConference[GROUP '23]{The 2023 ACM International Conference on Supporting Group Work (GROUP '23) Companion}{January 8--11, 2023}{Hilton Head, SC, USA}
\acmBooktitle{The 2023 ACM International Conference on Supporting Group Work (GROUP '23) Companion (GROUP '23), January 8--11, 2023, Hilton Head, SC, USA}
\acmDOI{10.1145/3565967.3571759}
\acmISBN{978-1-4503-9945-6/23/01}

\begin{document}

\title[Designing Human-Centered Algorithms for the Public Sector]{Designing Human-Centered Algorithms for the Public Sector \newline A Case Study of the U.S. Child-Welfare System}

\author{Devansh Saxena}
\affiliation{%
  \institution{Department of Computer Science, Marquette University}
  \city{Milwaukee}
  \state{Wisconsin}
  \country{USA}}
\email{devansh.saxena@marquette.edu}

\renewcommand{\shortauthors}{Devansh Saxena}

\begin{abstract}
The U.S. Child Welfare System (CWS) is increasingly seeking to emulate business models of the private sector centered in efficiency, cost reduction, and innovation through the adoption of algorithms. These data-driven systems purportedly improve decision-making, however, the public sector poses its own set of challenges with respect to the technical, theoretical, cultural, and societal implications of algorithmic decision-making. To fill these gaps, my dissertation comprises four studies that examine: 1) how caseworkers interact with algorithms in their day-to-day discretionary work, 2) the impact of algorithmic decision-making on the nature of practice, organization, and street-level decision-making, 3) how case notes can help unpack patterns of invisible labor and contextualize decision making processes, and 4) how casenotes can help uncover deeper systemic constraints and risk factors that are hard to quantify but directly impact families and street-level decision-making. My goal for this research is to investigate systemic disparities and design and develop algorithmic systems that are centered in the theory of practice and improve the quality of human discretionary work. These studies have provided actionable steps for human-centered algorithm design in the public sector.
\end{abstract}

\ccsdesc[500]{Human-centered computing~Human-computer interaction (HCI)}
\ccsdesc[300]{Human-centered computing~Empirical studies in HCI}
\ccsdesc[100]{Applied computing~Computing in government}

\keywords{algorithmic decision-making, discretion, bureaucracy, child-welfare system, computational narrative analysis}

\maketitle

\section{Introduction}
Over the past two decades, several high-stakes decision-making domains such as the child-welfare system (CWS), criminal justice system, education, and medical services have increasingly turned towards risk assessment algorithms as a means to standardize and improve decision-making. Facing severely limited resources and new dilemmas in the form of burdensome workloads and high staff turnover, most human services agencies have also turned towards algorithms as they purportedly promise to reduce costs and provide greater efficiencies in public policy and social services delivery. CWS has also been the center of public and media scrutiny because of the harm caused to children who are removed from the care of their parents \cite{camasso2013decision}. On the other hand, CWS also receives severe criticism and media attention for child abuse tragedies where the system failed to remove and protect a child \cite{gajanan_2020}. This has further mounted the pressure on CWS in several states in the United States (U.S.) to employ structured decision-making tools (and more recently, algorithmic decision-making) to prove that they are employing evidence-based, consistent, and objective decision-making processes \cite{saxena2020conducting, saxena2020child}. Decades of research in clinical psychology and medicine exhibit that statistical decision-making outperforms human experts in prediction tasks \cite{grove2000clinical, aegisdottir2006meta} and is often cited as a justification for introducing algorithms in the public sector. However, as illustrated by my CHI 2020 literature review \cite{saxena2020human}, CWS poses its own challenges with respect to the \textit{\textbf{technical}} (i.e., quality of data, reliability/validity of constructs), \textit{\textbf{social and cultural}} (i.e., workers' interactions with algorithms, impact of systemic constraints), \textit{\textbf{theoretical}} (i.e., what is empirical risk vs. theoretical risk?), and \textit{\textbf{societal}} (i.e., impact of algorithms on communities and decision-making ecosystem) implications of algorithmic decision-making.

\vspace{0.15cm}
Abebe et al. \cite{abebe2020roles} highlight that much of the computational research that focuses on fairness, bias, and accountability on algorithmic systems continues to formulate “fair” technical solutions while failing to address deeper systemic and structural injustices. Through my dissertation work, I bring attention back to the \textit{sociotechnical} and highlight social problems in child-welfare and how these problems become embedded in algorithmic systems. Through the studies discussed below, my dissertation assumes the dual roles of \textit{computing as rebuttal} where I highlight the technical limitations and feasibility of risk assessment algorithms, and of \textit{computing as synecdoche} by uncovering systemic complexities and social problems that directly impact families. This dissertation will also seek to make contributions at the intersection of gaps highlighted by the literature review and recommend solutions centered in strength- and asset-based approaches \cite{bronfenbrenner1975reality, zimmerman2013resiliency, badillo2018chibest} that will improve the state of current algorithmic interventions, enhance child-welfare practice, and improve street-level decisions mediated through algorithms. 

Therefore, my dissertation answers the following overarching research questions: 

\begin {itemize} [leftmargin=*]
  \item \textbf{RQ1:} (a) How do caseworkers interact with algorithms in their daily lives, and (b) How does the implementation of a given algorithm impact algorithmic decision-making, human discretion, and bureaucratic processes? 
  \item \textbf{RQ2:} (a) Can computational text analysis help uncover invisible patterns of human discretionary work conducted within the constraints of bureaucracy, and (b) can these theoretical signals derived from casenotes help contextualize algorithmic decisions?
  \item \textbf{RQ3:} \textit(a) How is "risk" quantified empirically within algorithmic systems as compared to how it is understood theoretically within the domain?, and (b) how do risk factors fluctuate and mediate each other throughout the child-welfare process and its implications for algorithmic decision-making?
\end{itemize}

To answer these questions, I will conduct the four studies described below. Examining the nature of practice and street-level discretionary work as well as the impact of systemic and policy-related barriers on decision-making (human or algorithmic) will allow us to develop technical solutions that operate within these constraints and augment the quality of human discretionary work.

\section{Research Overview}
In the following sections, I provide a short overview of my four dissertation studies.

\subsection{\large{Study 1: Theoretical Framework for Algorithmic Decision-Making in the Public Sector Developed through an Ethnography of Child-Welfare}}

This study constitutes an in-depth ethnographic case study that I conducted at a child-welfare agency in Milwaukee, Wisconsin \cite{saxena2021framework}. It was published at CSCW '2021 and was presented at the conference. It contributes to the \textbf{\textit{theoretical}} and \textbf{\textit{social and cultural}} gaps highlighted by the literature review. Algorithms in the public sector is a domain in its own right and requires a cohesive framework that explains how algorithms interact with bureaucracy and human discretion. First, drawing upon theories from Human-Computer Interaction (HCI), Science and Technology Studies (STS), and Public Administration (PA), we propose a theoretical framework for algorithmic decision-making for the public sector (ADMAPS) which accounts for the interdependencies between human discretion, bureaucratic processes, and algorithmic decision-making. The framework is then validated through a case study of algorithms in use at the agency. Second, the ethnography uncovers the daily algorithmic practices of caseworkers, what causes them to (dis)trust an algorithm, and how they navigate through different algorithmic systems especially when they do not account for policy and systemic barriers or resource constraints at the agency.

\subsection{\large{Study 2: Examining Invisible Patterns of Street-level Discretionary Work in Child Welfare embedded in Caseworker Narratives}}

This study seeks to utilize sources of information that have been hard to quantify so far, namely, caseworker narratives. Child-welfare caseworkers are trained in writing detailed casenotes about their interactions with families and case progress through the life of the case. This study contributes to the \textbf{\textit{technical}} and \textbf{\textit{theoretical}} gaps illustrated by the literature review by deriving rich qualitative signals from casenotes using natural language processing techniques such as topic modeling. We are specifically analyzing casenotes written by the Family Preservation Services (FPS) team that works closely with birth parents in their efforts to achieve reunification. Casenotes offer a rich description of decisions, relationships, conflicts, personas, as well as policy-related and systemic barriers. Analyzing these casenotes offers a unique lens towards understanding the workings of a child-welfare team trying to achieve reunification; one of the primary policy-mandated goals of CWS. Theoretical signals derived from casenotes will also help contextualize the quantitative structured assessments \cite{saxena2022train} and highlight patterns of invisible labor conducted by caseworkers and systemic constraints and power asymmetries that impact all decisions. This study was published at CHI'2022 \cite{saxena2022unpacking}.

\subsection{\large{Study 3: Algorithms in the Child-Welfare Ecosystem: Impact on Practice, Organization, and Street-Level Decision-Making}}

Drawing upon findings from a two-year ethnography conducted at a child-welfare agency, we highlight how algorithmic systems are embedded within a complex decision-making ecosystem at critical points of the child-welfare process. In our prior study \cite{saxena2021framework}, we focused on the micro-interactions between the dimensions of human discretion, algorithmic decision-making, and bureaucratic processes to understand why algorithms failed (or succeeded) to offer utility to child-welfare staff and their impact on the quality of human discretionary work. In this study, we critically investigate the macro-interactions between these three elements to assess the impact of algorithmic decision-making on the nature of practice, the organization, as well as the interactions between human discretion and bureaucratic processes to understand how the nature of street-level decision-making is changing and whether algorithms are living up to the promises of cost-effective, consistent, and fair decision-making. This study contributes to the \textbf{\textit{social and cultural}} and \textbf{\textit{societal}} gaps highlighted by the literature review by unpacking how the decision-making ecosystem within the public sector is changing. It also depicts the case study of an algorithm that offers higher utility to caseworkers, however, required significant investments from the agency leadership to bring about that ecological change in decision-making where the algorithmic system plays an essential role. This manuscript is currently under review for the ACM Journal on Responsible Computing in October 2022.

\subsection{\large{Study 4: Rethinking "Risk" in Algorithmic Systems Through A Computational Narrative Analysis of Casenotes in Child-Welfare}}

Risk assessment algorithms have been adopted by several public sector agencies to make high-stakes decisions about human lives. However, there is a mismatch between how risk is quantified empirically based on administrative data versus how it is understood theoretically within the domain. Public servants such as caseworkers are essentially risk workers who are tasked with assessing and managing risks, translating risk in different contexts, and conducting care work in the context of risk \cite{gale2016towards}. However, this risk work is increasingly mediated through algorithmic systems with a mismatch between \textit{empirical risk} and \textit{theoretical risk} that leads to unreliable decision-making and conflicts in practice. This study contributes to the \textbf{\textit{theoretical}} and \textbf{\textit{societal}} gaps highlighted by the literature review. Algorithms model “risk” based on individual client characteristics to identify clients most in need. However, this understanding of risk is primarily based on easily quantifiable risk factors that present an incomplete and biased perspective of clients. In this study, I conducted computational narrative analysis of child-welfare casenotes and draw attention toward deeper systemic risk factors that are hard to quantify but directly impact families and street-level decision-making. Beyond individual risk factors, the system itself poses a significant amount of risk to families where parents are over-surveilled by caseworkers and experience a lack of agency in decision-making. I also problematize the notion of risk as a static construct by highlighting temporality and mediating effects of different risk and protective factors and show that any temporal point estimate of risk will produce biased predictions. I also draw caution against using casenotes in NLP-based algorithms by unpacking their limitations and biased embedded within them. This study is currently under submission at CHI'2023.

\vspace{0.2cm}
\section{Research Progress \& GROUP 2022 DC Participation}
All four studies have been completed with \textbf{Study 1} and \textbf{Study 2} published and presented at their respective conferences. The manuscript for \textbf{Study 3} is currently under submission and review for the ACM Journal on Responsible Computing. \textbf{Study 4} is currently under submission and review at CHI'2023.  

\vspace{0.2cm}
\section{EXPECTED OUTCOMES}
My dissertation assumes the dual roles of \textit{computing as rebuttal} and \textit{computing as synecdoche} and will make three contributions. First, I highlight the technical limitations and feasibility of risk assessment algorithms and draw attention to the systemic complexities and structural issues that directly impact families. Second, I developed a theoretical framework for algorithmic decision-making in the public sector that accounts for the complex interdependencies between human discretion, bureaucratic processes, and algorithmic decision-making. Third, I show how computational narrative analysis can help uncover patterns of invisible labor, systemic constraints, and power asymmetries and problematize the empirical notion of risk by highlighting the temporality of risk as well as systemic risk factors that are hard to quantify but directly impact street-level decision-making. 

\bibliographystyle{ACM-Reference-Format}
\bibliography{bibliography}

\end{document}